\documentclass[12pt]{article}
\def\cc{\c{c}}

\def\ca{\c{c}\~{a}}

\def\ii{\'{\i}}

\usepackage{epsfig}

\begin{document}

{\centerline{\bf Effective Lagrangian approach to multi-quark interactions}}
\vspace{0.5cm}

{\centerline{A. A. Osipov, B. Hiller, A. H. Blin}}
\vspace{0.5cm}

\noindent{\it{Centro de F\'{\i}sica Computacional, Departamento de
F\'{\i}sica da Universidade de Coimbra,}}
\centerline{ 3004-516 Coimbra, Portugal}

\vspace{0.5cm}

\begin{abstract}
In this workshop we have presented the results obtained in the three-flavour ($N_f=3$)
Nambu--Jona-Lasinio model Lagrangian which includes all non-derivative vertices at NLO in the $1/N_c$ expansion
of spin zero multi-quark interactions. In particular the role played by the explicit chiral symmetry breaking interactions has been discussed in comparison with previous model Lagrangians.  
\end{abstract}

The subject of this year's Bled workshop is "Quark masses and hadron spectra". The understanding of the origin of masses from fundamental principles may have moved a step closer with the announcement of the existence of the Higgs, however the reason for the hierarchy of masses observed for several families of leptons and quarks still eludes us. The current quark masses are external to the gauge principle underlying the foundations of QCD. In an effective approach to QCD the most innocuous way is to consider them born from external sources interacting with originally massless fields which comply with all the symmetries. If in addition the study of strong interactions is limited to  the energy range which is
of order $\Lambda\simeq 4\pi f_\pi\sim 1$\ GeV \cite{Georgi:1984}, where $\Lambda$ 
characterizes the scale for spontaneous chiral symmetry $\chi_S$ breaking, a firm set-up for its systematic inclusion is supplied by the seminal papers of Nambu and Jona-Lasinio (NJL) \cite{Nambu:1961}.

 Our procedure relies on the very general assumption that this scale determines the hierarchy of local multi-quark interactions which model QCD at low energies. It has been pointed out in \cite{Andrianov:1993a,Andrianov:1993b} that it is sufficient to truncate the tower of multi-quark interactions at 8 quarks (q)  to complete in $4D$ the number of vertices relevant at the scale of dynamical chiral symmetry breaking.

The $U(1)_A$ symmetry breaking 't Hooft $(2 N_f)$ flavor determinant \cite{Hooft:1976,Hooft:1978} adds $1/N_c$ suppressed interactions to the original NJL Lagrangian \cite{Bernard:1988,Reinhardt:1988}. Having  first focussed on the resolution of the instability of this model's effective potential \cite{Osipov:2006b}, we have enlarged the Lagrangian by a general set of equally suppressed spin zero 8q interactions  \cite{Osipov:2005b,Osipov:2006a}. 

Later on, showing that the  $N_c$ counting rules are congruent with the classification of vertices in terms of the $\chi_S$ breaking scale, we have taken into consideration the terms of higher order in the current quark-mass expansion \cite{Osipov:2013a,Osipov:2013b}, which are responsible for the explicit
chiral symmetry breaking at the same order as the 't Hooft determinant and eight quark terms
previously analyzed. The standard mass term of the free
Lagrangian is only a part of the more complicated picture arising in effective
models beyond leading order \cite{Gasser:1982}. Chiral perturbation theory
\cite{Weinberg:1979,Pagels:1975,Gasser:1984}
gives a well-known example of a self consistent accounting of the mass terms, order by order, in
an expansion in the masses themselves. 

Using path integral bosonization techniques which take
appropriately into account the quark mass differences \cite{Osipov:2001a,Osipov:2001b}, the mesonic
Lagrangian up to three-point mesonic vertices is obtained in \cite{Osipov:2013b}. 

We end up with $4+10=14$ low-energy constants at leading and NLO of the effective $1/N_c$ expansion.  The  model parameters are fully controlled on the theoretical side by symmetry arguments and on the experimental side by the characteristics of the low lying pseudoscalars and scalars. The number of observables described until now by far surpasses  the number of parameters \cite{Osipov:2013b}.  

The tree level bosonized Lagrangian carries either signatures of violation of the Zweig-rule or of
admixtures of $q^2{\bar q}^2$ to the quark-antiquark states.
Elsewhere these are obtained by considering explicitly meson loop corrections, tetraquark
configurations and so on \cite{Jaffe:1977,Black:1999,Wong:1980,Narrison:1986,Beveren:1986, Latorre:1985,Alford:1998,Achasov:1984,Isgur:1990,Schechter:2008,
Close:2002,Klempt:2007}.

By calculating the mass
spectra and the strong decays of the scalars, one can realize which multi-quark
interactions are most relevant at the scale of spontaneous $\chi_S$ breaking. On the other
hand, by analyzing the two photon radiative decays, where a different scale,
associated with the electromagnetic interaction, comes into play, one can study
the possible recombinations of quarks inside the hadron. 

Our main results are so far: 

1-We achieve 
total agreement with the empirical low lying pseudoscalar
meson spectrum. The current quark mass dependent interaction terms mainly
responsible for the fit belong to the class of OZI-violating interactions,  
representing additional corrections to the 't Hooft $U_A(1)$ breaking
mechanism.  Explicit $\chi_S$ breaking effects in interactions are essential to obtain the 
empirical ordering $m_K < m_\eta$ and the magnitude of the splitting. 
The fit of the $\eta\!-\!\eta'$ mass splitting together with the overall
successful description of the whole set of low-energy pseudoscalar
characteristics is actually a solution for a long standing problem of NJL-type
models.  

2-We are also capable to describe the spectrum of the light scalar nonet. 
The explicit $\chi_S$ breaking terms related with $q^2{\bar q}^2$ admixtures are the main source of the fit associated with the empirical ordering $m_{\kappa_0} < m_{a_0}\sim m_{f_0}$. 
On the other hand, the mixing angle of the singlet-octet scalar states $\theta_S$ as well as the mass of the $\sigma$ meson
are strongly affected by OZI-violating short range forces.
 
3-With all parameters of the model fixed by the spectra we
analyzed a bulk of two body decays at tree level of the bosonic
Lagrangian: the strong decays of the scalars $\sigma\rightarrow\pi\pi$,
$f_0(980)\rightarrow\pi\pi$, $\kappa(800)\rightarrow\pi K$, $a_0(980)
\rightarrow\pi\eta$, as well as the two photon decays of $a_0(980)$, $f_0(980)$
and $\sigma$ mesons and the anomalous decays of the pseudoscalars $\pi
\rightarrow\gamma\gamma$, $\eta\rightarrow\gamma\gamma$ and $\eta'\rightarrow
\gamma\gamma$. 

Our results for the strong decays of the scalars are within the current
expectations.
The radiative decays of the scalars are smaller than the observed
values for the $f_0(980)$ and the $\sigma$, but reasonable for the $a_0$.

We obtain that the $a_0(980)$ meson couples with a large strength of the
multi-quark components to the two kaon channel in its strong decay to two
pions, but evidences a dominant $q\bar q$ component in its radiative decay.
As opposed to this, the $\sigma$ and $f_0(980)$ mesons do not display an enhanced
$q\bar q$ component neither in their two photon decays nor in the strong
decays. The widths of the $a_0(980)\to\pi\eta$ and $f_0(980)\to\pi\pi$ decays are well
accomodated within a Flatt\'e description. We corroborate other model calculations
in which the coupling of the $f_0(980)$ and $a_0(980)$ mesons to the $K \bar K$
channel is needed for the description of the decays $f_0(980)\rightarrow\pi\pi$
and $a_0(980)\rightarrow\pi\eta$. We find that this coupling is most crucial
for the latter process. 

The radiative decays of the scalar mesons into two photons show that the main
channel for the $a_0(980)$ decay proceeds through coupling to a $q\bar q$
state, while the radiative decays of singlet-octet states $\sigma, f_0$ must
proceed through more complex strutures. This does not mean that the $a_0$ meson is mainly a $q \bar q$ state,
but that the multi-quark component with the large strength in the two kaon
channel, important for the reduction of the $a_0 \pi\eta$ strong decay width,
is not the leading process in the two photon decay of this meson.

Finally, the radiative decays of the pseudoscalars are in very good
agreement with data.

4-
The response to the external parameters $T,\mu$ has been recently addressed in \cite{Moreira:2014}, with implications on strange quark matter formation. 
In the early version of the model Lagrangian without the explicit NLO $\chi_S$ breaking terms one obtains that although the vacuum properties remain almost insensitive to the 8q interactions, their effects are remarkable for medium and thermal properties \cite{Osipov:2007b,Osipov:2007c,Kashiwa:2007,Kashiwa:2008,Hiller:2010,Bhattacharyya:2010}, as well as in presence of a strong constant magnetic field  \cite{Osipov:2007a,Gatto:2010,Gatto:2011,Frasca:2011}.

\vspace{0.5cm}

{\bf Acknowledgements}
Research supported by Centro de F\ii sica Computacional of the University of Coimbra, by Funda\ca o para a Ci\^encia e Tecnologia, Or\cc amento de Estado
and by the European Community-Research Infrastructure Integrating Activity Study of Strongly Interacting Matter (Grant Agreement 283286) under
the Seventh Framework Programme of EU.



\begin{thebibliography}{99}

\bibitem{Georgi:1984} A. Manohar, H. Georgi, Nucl. Phys. B \textbf{234}, 189
   (1984).
\bibitem{Nambu:1961} Y. Nambu, G. Jona-Lasinio,
   Phys. Rev. \textbf{122}, 345 (1961); \textbf{124}, 246 (1961).
\bibitem{Andrianov:1993a} A. A. Andrianov, V. A. Andrianov, Theor. Math.
   Phys. \textbf{94}, 3 (1993).
\bibitem{Andrianov:1993b} A. A. Andrianov, V. A. Andrianov, Int. J. of Mod.
   Phys. A \textbf{8}, 1981 (1993).
\bibitem{Hooft:1976} G. 't Hooft, Phys. Rev. D \textbf{14}, 3432 (1976).
\bibitem{Hooft:1978} G. 't Hooft, Phys. Rev. D \textbf{18}, 2199 (1978).
\bibitem{Bernard:1988} V. Bernard, R. L. Jaffe, U.-G. Mei\ss ner,
   Phys. Lett. B \textbf{198}, 92 (1987).
\bibitem{Reinhardt:1988} H. Reinhardt and R. Alkofer,
   Phys. Lett. B \textbf{207}, 482 (1988).
\bibitem{Osipov:2006b} A. A. Osipov, B. Hiller, V. Bernard, A. H. Blin,
Ann. of Phys. \textbf{321}, 2504 (2006). 
\bibitem{Osipov:2005b} A. A. Osipov, B. Hiller, J. da Provid\^encia, Phys.
   Lett. B \textbf{634}, 48 (2006).
\bibitem{Osipov:2006a} A. A. Osipov, B. Hiller, A. H. Blin,
   J. da Provid\^encia, Ann. of Phys. \textbf{322}, 2021 (2007).
\bibitem{Osipov:2013a} A. A. Osipov, B. Hiller, A. H. Blin, Eur. Phys. J. A
   {\bf 49}, 14 (2013).
\bibitem{Osipov:2013b} A. A. Osipov, B. Hiller, A. H. Blin, Phys. Rev. D 
   {\bf 88},  054032  (2013).
\bibitem{Gasser:1982} J. Gasser, H. Leutwyler, Phys. Rep. \textbf{87}, 77
   (1982).
\bibitem{Weinberg:1979} S. Weinberg, Physica A \textbf{96}, 327 (1979).
\bibitem{Pagels:1975} H. Pagels, Phys. Rev. C \textbf{16}, 219 (1975).
\bibitem{Gasser:1984} J. Gasser, H. Leutwyler, Ann. of Phys. \textbf{158},
   142 (1984).
\bibitem{Osipov:2001a} A. A. Osipov, B. Hiller, Phys. Lett. B \textbf{515}, 458 (2001).
\bibitem{Osipov:2001b} A. A. Osipov, B. Hiller, Phys. Rev. D \textbf{64}, 087701 (2001).   
\bibitem{Jaffe:1977} R. J. Jaffe, Phys. Rev. D \textbf{15}, 267 (1977);
   R. J. Jaffe, Phys. Rev. D \textbf{15}, 281 (1977).
\bibitem{Black:1999} D. Black, A. H. Fariborz, E. Sannino, J. Schechter, Phys. Rev. D \textbf{59},
   074026 (1999).
\bibitem{Wong:1980} D. Wong, K. F. Liu, Phys. Rev. D {\bf 21}, 2039 (1980)
\bibitem{Narrison:1986} S. Narrison, Phys. Lett. B {\bf 175}, 88 (1986).
\bibitem{Beveren:1986} E. van Beveren, T. A. Rijken, K. Metzger, C. Dullemond,
   G. Rupp, J. E. Ribeiro, Zeit. Phys. C \textbf{30}, 615 (1986).
\bibitem{Latorre:1985} J.I. Latorre, P. Pascoal, J. Phys. G {\bf 11}, L231
   (1985)
\bibitem{Alford:1998} M. Alford, R. L. Jaffe, Nucl. Phys. B {\bf 509}, 312
   (1998)
\bibitem{Achasov:1984} N. N. Achasov, S. A. Devyanin, D. N. Shestakov, Z. Phys.
   C {\bf 16}, 55 (1984)
\bibitem{Isgur:1990} N. Isgur, J. Weinstein, Phys. Rev. D {\bf 41}, 2236 (1990).
\bibitem{Schechter:2008} A. H. Fariborz, R. Jora, J. Schechter, Phys. Rev. D
   \textbf{77}, 094004 (2008).
\bibitem{Close:2002} F. E. Close, N. A. T\"ornqvist, J. Phys. G: Nucl. Part.
   Phys. \textbf{28}, R249 (2002). 
\bibitem{Klempt:2007} E. Klempt, A. Zaitsev, Phys. Rep. \textbf{454}, 1 (2007).
\bibitem{Moreira:2014} J. Moreira, J. Morais, B. Hiller, A.A. Osipov, A.H. Blin, arXiv:1409.0336 [hep-ph]
\bibitem{Osipov:2007b} A. A. Osipov, B. Hiller, J. Moreira, A.H. Blin, J. da Providencia, Phys. Lett. B \textbf{646}, 91 (2007).  
\bibitem{Osipov:2007c} A. A. Osipov, B. Hiller, J. Moreira, A.H. Blin,  Phys. Lett. B \textbf{659}, 270 (2008). 
\bibitem{Kashiwa:2007} K. Kashiwa, H. Kouno, T. Sakaguchi, M. Matsuzaki, M. Yahiro, Phys. Lett. B \textbf{647}, 446 (2007);
\bibitem{Kashiwa:2008} K. Kashiwa, H. Kouno, M. Matsuzaki, M. Yahiro, Phys. Lett. B 662, 26 (2008).
\bibitem{Hiller:2010} B. Hiller, J. Moreira, A. A. Osipov, A.H. Blin,  Phys. Rev. D \textbf{81},116005 (2010).
\bibitem{Bhattacharyya:2010} A. Bhattacharyya, P. Deb, S. K. Ghosh, R. Ray, Phys. Rev. D \textbf{82} (2010) 014021 
\bibitem{Osipov:2007a} A. A. Osipov, B. Hiller, A. H. Blin, J. da Provid\^encia,
   Phys. Lett. B \textbf{650} 262 (2007).
\bibitem{Gatto:2010} R. Gatto, M. Ruggieri, Phys. Rev. D \textbf{82}, 054027
   (2010).
\bibitem{Gatto:2011} R. Gatto, M. Ruggieri, Phys. Rev. D \textbf{83}, 034016
   (2011).
\bibitem{Frasca:2011} M. Frasca, M. Ruggieri, Phys. Rev. D \textbf{83},
   094024 (2011).
\end{thebibliography}
\end{document}